\documentclass{article}
\usepackage{spconf,amsmath,graphicx,url}
\usepackage[bookmarks=false]{hyperref}

\title{High Resolution Guitar Transcription via Domain Adaptation}
\name{Xavier Riley*, Drew Edwards*, Simon Dixon\thanks{*Equal contribution. \\ XR and DE are research students at the UKRI Centre for Doctoral Training in Artificial Intelligence and Music. XR is supported by UK Research and Innovation [grant number EP/S022694/1]; DE is supported by Queen Mary University of London and Yamaha.}}
\address{Centre for Digital Music, Queen Mary University of London}
\begin{document}
%
\maketitle
\begin{abstract}
Automatic music transcription (AMT) has achieved high accuracy for piano due to the availability of large, high-quality datasets such as MAESTRO and MAPS, but comparable datasets are not yet available for other instruments. In recent work, however, it has been demonstrated that aligning scores to transcription model activations can produce high quality AMT training data for instruments other than piano.
Focusing on the guitar, we refine this approach to training on score data using a dataset of commercially available score-audio pairs. We propose the use of a high-resolution piano transcription model to train a new guitar transcription model. The resulting model obtains state-of-the-art transcription results on GuitarSet in a zero-shot context, improving on previously published methods.
\end{abstract}
\section{Introduction}
\label{sec:intro}

Automatic Music Transcription (AMT) is a widely studied music information retrieval (MIR) task. Recent work has shown a progression of good results for piano transcription in particular~\cite{oaf, kong, seq2seqPiano}, arising from applying deep learning techniques to large, high quality datasets such as MAESTRO~\cite{maestro} and MAPS~\cite{maps}. While this work indicates that accurate AMT is possible in principle, achieving a similar level of performance for instruments beyond piano remains challenging, which is in part due to a lack of high quality training data~\cite{mt3}.

Taking guitar as an example, it is possible to obtain accurate ground truth data by means of specialist equipment such as individual string (hexaphonic) pickups~\cite{guitarset}. However, this is very costly in terms of time, since there is no existing large dataset, and new recordings would need to be made. Furthermore, hexaphonic pickups do not provide perfect separation (there is some bleed from adjacent strings), and the resulting signals still need to be transcribed and checked.

The MusicNet dataset~\cite{thickstun2017learning} attempted to address the lack of data via automatic score alignment (ASA), using dynamic time warping (DTW) to align score data to spectral features from audio. A new approach was put forward by Maman and Bermano \cite{maman} in which activations from an existing transcription model are used instead of spectral features. In their work this model is bootstrapped using synthetic audio-score pairs and then retrained on the aligned scores. This process goes through several iterations of Expectation Maximisation \cite{EM} to improve the results before a final transcription model is obtained. They report strong results for transcription of GuitarSet~\cite{guitarset} in a zero-shot setting, suggesting that the approach generalises well.


Outside of GuitarSet, there is a lack of aligned annotations for guitar due to the difficulties of data collection and creation. For this reason, existing guitar transcription approaches make use of GuitarSet with a train/test split described in \cite{mt3}.

\section{Method}
\label{sec:format}


We begin by describing our choice of model and the initial data augmentation pretraining stage. We then discuss the data sources we used. Following Maman and Bermano, who showed that automatic score alignment can be sufficiently accurate to produce useful training data~\cite{maman}, we then describe the steps of aligning the audio to scores using an adaptation of their method, highlighting any relevant differences to their work. Finally, we fine-tune our model using the new dataset and report results on GuitarSet and a test split of our new dataset.

\subsection{Choice of Model and Augmentation}
\label{ssec:model_choice}


A key difference in this work is in the choice of the underlying transcription model used both for the alignment process and the fine-tuning.  Given that any score-audio alignment method will have some non-negligible error, it is reasonable to expect a degree of misalignment between note-onsets and their labels. Therefore, a model that is more robust to noisy training data is highly desirable. We use the High-Resolution Piano Transcription model by Kong et al.\ \cite{kong}, while Maman and Bermano use the Onsets and Frames model \cite{oaf}. Kong et al. demonstrate that their high-resolution approach is better than Onsets and Frames when it comes to data with misaligned labels: they train both models with 50 millisecond misalignments, and their model yields a 96.49\% note-onset F1-score, while Onsets and Frames falls to 76.52\% \cite[Table III]{kong}.

We first re-trained the High-Resolution model on the MAESTRO piano dataset using the data augmentation described in \cite{maestro} (pitch shift, EQ, compression, reverb, and noise), along with random transpositions of up to $\pm2$ semitones. This re-trained model performs better than the published model (trained without data augmentation) on out-of-distribution samples, such as the MAPS dataset (88.1\% vs. 82.4\% note-onset F1-score on MAPS test set).

\subsection{Score and Audio Data Sources}

Our work also differs from Maman and Bermano as follows. Their pipeline allows for scores to be aligned in an unsupervised fashion. They achieve this by rejecting sections of score-audio pairs where either a) the DTW alignment process collapses multiple source timesteps onto a single target timestep (``singular points'') or b) where the model activations for an aligned region are below a given threshold. As we are using a dataset of professionally transcribed scores with good correspondence between audio and MIDI, we find that the initial alignment is ready to use as there are no significant errors. This simplifies the implementation and helps to demonstrate that a smaller number of high quality audio-score pairs is sufficient to train a successful model.

For the score-audio pairs we use a set of 79 professionally transcribed scores which focus on solo guitar in the jazz genre. The data, totalling 4 hours of audio, covers 33 guitarists from recordings over the last 65 years playing various types of guitars (steel string, nylon string, electric, acoustic) in a range of recording conditions. 
The transcriptions are commercially available\footnote{\url{https://www.francoisleduconlinelibrary.com/}} and the corresponding audio for each specific transcription is available on YouTube. Scores were provided in GuitarPro format, and were converted to MIDI using MuseScore3. We emphasise that our  method is not reliant on this specific data, and that the production of a large scale, freely licensed dataset for guitar is a target of future work. A complete list of track titles and split groupings can be found in the supplementary material\footnote{\url{https://xavriley.github.io/HighResolutionGuitarTranscription/}}. The aligned MIDI files will also be available from the authors on request.

%
%

\subsection{Audio-Score Alignment}
\label{ssec:audio_score_alignment}

Following Maman and Bermano, we use a two-stage alignment process with some variations on their method. The first stage (``coarse alignment'') obtains a mapping between transcription model activations and the score (represented by MIDI) using DTW. Instead of using synthetic data to bootstrap the transcription model, we start with our pre-trained High-Resolution piano transcription model described in \ref{ssec:model_choice}. The output of the initial alignment stage is an alignment path which maps score timing to the timeline of the audio file.

The second stage (``fine alignment'') resolves any asynchronies in the performance of chords. In the score these are simultaneous in time, however in performance it is rare for pitches to be sounded at exactly the same time. We then align each note onset to its local maximum within the model activations \cite[appendix A.1.2]{maman}.

\begin{figure}
 \centerline{\framebox{
 \includegraphics[width=0.9\columnwidth]{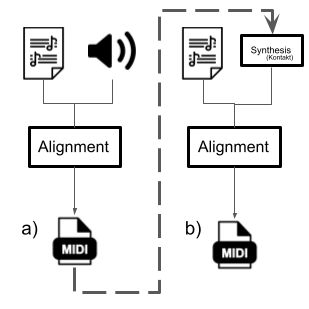}}}
 \caption{Diagram of the process used to validate the alignment accuracy of our proposed method}
 \label{fig:resynthesis}
\end{figure}

For this work we validate the alignment accuracy using synthesised guitar audio. Using our test split, we obtain performance aligned MIDI using the method described above (see Figure \ref{fig:resynthesis}a). We then re-synthesize this aligned MIDI using a high quality guitar sample library\footnote{\url{https://impactsoundworks.com/product/shreddage-3-archtop/}}, taking care to ensure that this process does not alter the timing of the original MIDI input.
With this newly synthesised audio, we repeat the alignment process using the original unaligned score MIDI to obtain a new prediction (see Figure \ref{fig:resynthesis}b). This is compared with the original alignment using the note-onset F-measure (no offset) metric, as implemented in the \texttt{mir\_eval} library~\cite{mireval}.

\subsection{Fine-Tuning Transcription Models}

Having created a high-precision dataset of aligned guitar transcriptions, we now apply this data to the downstream task of training an automatic guitar transcription model. We approach this task as a domain adaptation task, where we take an existing state-of-the-art piano transcription model, trained on a large dataset, and fine-tune it using our much smaller guitar transcription dataset. The source model we use is the same High-Resolution Piano Transcription model by Kong et al.\ \cite{kong}, described in Section \ref{ssec:model_choice}.

We take an 80-10-10 split (by pieces) of our guitar transcription dataset. For each recording, we create 10-second samples with a hop-size of 1 second (so there is overlap amongst the training samples). We experiment with fine-tuning using a variety of learning rates ($10^{\{-3, -4, -5\}}$) and use a batch size of $4$. Given the smaller size of our dataset, we retain the random transposition augmentation. All experiments run for 100K steps, approximately 10 epochs, and scale the learning rate by $0.9$ every 10K steps. Most experiments had very similar test scores and began converging within 30K steps. We selected the model with a learning rate of $10^{-5}$, where the difference between train and validation performance was minimized.

\section{Results}

\subsection{Validating Alignment Accuracy}

Results for validating the alignment process on our test subset are shown in Table \ref{tab:alignment}. Here we can see that the proposed alignment algorithm is able to recover polyphonic onsets with high precision and reliability, as compared with the alignment obtained from the initial DTW pass alone. Although synthetic data is likely to be easier to align (e.g.\ due to lack of reverberation), this shows that to obtain precise onset timing, coarse alignment using DTW alone is insufficient.

\begin{table}
 \begin{center}
 \begin{tabular}{|c||c|c|}
  \hline
  Threshold & Coarse (DTW-only) & Fine \\
  \hline
  50ms & 62\% & 98\% \\
  25ms & 27\% & 95\% \\
  \hline
 \end{tabular}
\end{center}
 \caption{Results for note-level Onset-only F1-measure when realigning a score to synthesised guitar audio}
 \label{tab:alignment}
\end{table}

\subsection{Guitar Transcription}

We report transcription results for three settings: the GuitarSet test split (supervised), the entire GuitarSet (zero-shot) and our own test split, which is a withheld test subset of the data obtained by the proposed alignment method.

Table \ref{tab:guitarset_sup} shows results for other models in a supervised setting where GuitarSet data was seen in the training stage. Our method does not improve on the best reported results, however we are within 2\% of the highest performing model for the onset-only F-measure.

\begin{table}
 \begin{center}
 \begin{tabular}{|l|l|l|l|}
  \hline
  & $P_{50}$ & $R_{50}$ & $F_{50}$ \\
  \hline
  Basic Pitch \cite{bittner} & - & - & 79.0 \\
  Omnizart \cite{bittner} & - & - & 59.0 \\
  Kim et al. \cite{kim} & 78.1 & 77.7 & 77.5 \\
  MT3 \cite{mt3} & - & - & 90.0 \\
  TabCNN (in \cite{fretnet}) & 54.8 & 65.6 & 58.3 \\
  FretNet \cite{fretnet} & 90.9 & 54.5 & 66.4 \\
  Lu et al. \cite{tfperceiver} & - & - & \textbf{91.1} \\
  SpecTNT (in \cite{tfperceiver}) & - & - & 90.7 \\
  Ours ($_\mathit{FL}$) & 87.6 & 86.8 & 86.9 \\
  Ours ($_\mathit{GS+FL}$) & \textbf{91.1} & \textbf{88.5} & 89.7 \\ 
  \hline
  
 \end{tabular}
\end{center}
 \caption{Results for supervised note-level transcription accuracy on the GuitarSet test split. $P_{50}$, $R_{50}$, and $F_{50}$ are Precision, Recall and Onset-only F1-measure, expressed as percentages, at 50ms resolution. All are evaluated on onsets only (no offsets or velocity), using the \texttt{mir\_eval} library. FL refers to our dataset (François Leduc) and GS refers to GuitarSet where these were used as training data.}
 \label{tab:guitarset_sup}
\end{table}

To put these results in context however, we compare them with the published results for GuitarSet in the zero-shot setting (Table \ref{tab:guitarset_zs}). Here we see that our model and dataset are able to perform at close to supervised levels across the entire dataset, improving by over 4\% on the next best method (Maman and Bermano). Transcription evaluation was performed on the entire dataset for all models.

\begin{table}
 \begin{center}
 \begin{tabular}{|l|l|l|l|}
  \hline
  & $P_{50}$ & $R_{50}$ & $F_{50}$ \\
  \hline
  MT3 \cite{mt3} & - & - & 32.0 \\
  Kong et al.~\cite{kong} & 67.5 & 49.7 & 54.8 \\
  Kong et al.\ (w/ aug) & 80.6 & 44.0 & 50.3 \\
  Maman (MusicNet$_\mathit{EM}$) \cite{maman} & 86.6 & 80.4 & 82.9 \\
  Maman (Guitar) \cite{maman} & 86.7 & 79.7 & 82.2 \\
  Our approach ($_\mathit{FL}$) & \textbf{88.0} & \textbf{87.1} & \textbf{87.3} \\
  \hline
  
 \end{tabular}
\end{center}
 \caption{Results for note-level transcription accuracy on the entire GuitarSet in the zero-shot setting. For description of abbreviations see Table \ref{tab:guitarset_sup}.}
 \label{tab:guitarset_zs}
\end{table}

This shows that our approach generalises more consistently to guitar recordings than previously published methods. To reinforce this, we include results for our own test split, made up of score alignments for a diverse set of 9 commercial jazz guitar recordings. Again, our method shows superior performance and improves on Lu et al.\ by over 4\%, despite their method being the state-of-the-art for GuitarSet in the supervised setting. Results are shown in Table \ref{tab:beyondpiano}.

\begin{table}
 \begin{center}
 \begin{tabular}{|l|l|l|l|}
  \hline
   & $P_{50}$ & $R_{50}$ & $F_{50}$\\
  \hline
  Basic Pitch \cite{bittner} & 54.6 & 85.0 & 66.1 \\
  Omnizart \cite{omnizart} & 63.0 & 72.1 & 67.1 \\
  MT3 \cite{mt3} & 48.8 & 57.0 & 52.4 \\
  Lu et al. \cite{tfperceiver} & 83.6 & 77.3 & 80.0 \\
  Kong et al.~\cite{kong} & 59.4 & 59.5 & 52.0 \\
  Kong et al.~(aug) & 70.9 & 63.9 & 65.6 \\
  Our approach ($_\mathit{GS+FL}$) & \textbf{83.9} & \textbf{85.5} & \textbf{84.7}\\
  \hline
 \end{tabular}
\end{center}
 \caption{Results for note-level transcription accuracy on a test split of our proposed dataset. For description of abbreviations see Table \ref{tab:guitarset_sup}.}
\label{tab:beyondpiano}
\end{table}

\section{Discussion}

Our results show state-of-the-art performance for our model across both our test set and the entire GuitarSet in a zero-shot context, along with competitive results on GuitarSet in the supervised context. The results on our test split (Table \ref{tab:beyondpiano}) are particularly encouraging as the audio is taken from commercial recordings with varying acoustic conditions and audio quality. This indicates that the proposed method may perform similarly on real world examples. To demonstrate this, we include in the supplementary material some example videos of real world guitar recordings\footnote{\url{https://xavriley.github.io/HighResolutionGuitarTranscription/}}. To be fair, we note that our system is specifically trained for guitar, while other systems are trained only for piano or for multiple instruments; but the point here is what is possible to achieve via domain adaptation with a relatively small amount of data.

For GuitarSet, our method was also sufficiently accurate to identify two files with errors in the ground truth alignments (see supplementary material for details). This further demonstrates the ongoing challenge of obtaining clean labels for transcription tasks as discussed in ~\cite{mt3}. Given the small difference we observe when training with our dataset and GuitarSet combined, this indicates that our proposed dataset is of comparable accuracy but it has the added advantage of a diverse range of timbres.

Reinforcing the findings of Maman and Bermano, we confirm that transcription data from human-transcribed scores can be sufficiently accurate to train a successful model. This may be partly a testament to the skill of the original transcriber, but also supports the idea that digitised scores are a useful source of data for advancing research on this task.

In comparing the performance of other models on our test set, we see that Basic Pitch, Omnizart and the augmented Kong model perform reasonably well. Given that the Basic Pitch model architecture is broadly similar to the Kong transcription model, we suspect that the performance of Basic Pitch could be improved further by adopting some of the onset post-processing techniques introduced by Kong et al.\ \cite{kong}, or by adopting a more comprehensive data augmentation strategy. The results from Lu et al.\ \cite{tfperceiver} are also very strong and show that this architecture has good potential to generalise, given appropriately diverse training data.

Regarding the relatively poor performance of MT3~\cite{mt3} on our proposed test set, MT3 attempts to predict notes for multiple instruments simultaneously which is a more difficult task. We report results using a flattened output (merging pitched instrument outputs, as per \cite{mt3}) and also removed overlapping notes (those completely covered by another note at the same pitch) to avoid skewing the evaluation. Despite this, the large difference between MT3's results on our test split and on GuitarSet (supervised) show that the model does not generalise well to unseen data, which is also demonstrated in the poor zero-shot results for GuitarSet.

Furthermore, we note the parameter counts of the models compared: Basic Pitch uses 17K, Omnizart uses 8M~\cite{wu_2020}, the Kong derived models (including ours) use around 20M~\cite{bittner, kong}, Lu et al.\ uses 35M and MT3 uses 77M~\cite{mt3}. This suggests that our proposed solution is not relying on an outsized model capacity to achieve its performance.

There are many aspects of guitar transcription that are not addressed in this work and are left for future work. As the data comes from jazz, it features mostly ``clean'' guitar sounds, without distortion or strong audio effects. We do not deal with extended playing techniques (pitch bends, harmonics, slides). However we only used a relatively small dataset to adapt the transcription system to guitar, so we expect that given suitable transcriptions it will be possible to extend this work to cover a wider range of guitar music.

Finally, we note that the model described here performs well on out of distribution samples from other instruments, with additional examples included in the supplementary data. Some of these are from plucked string instruments (mandolin, harp) but also include bowed strings such as the violin. As we do not have aligned scores for these instruments we are unable to provide numeric results, however this suggests that our method can be adapted to other instrument groups. Progress in this direction will depend on the availability of digitized score and audio pairs.

\section{Conclusion}

To address the problem of a lack of AMT training data for instruments other than piano, we adapt a recently proposed method for polyphonic score alignment to meet this need. We improve on the proposed method by replacing Onsets and Frames with a high-resolution piano transcription model, which yields greater accuracy due to an improved tolerance for misaligned labels. We also release our dataset of 79 solo guitar annotations, constructed using an accurate score alignment method.

 Our combination of dataset and model choice achieves a large increase in transcription accuracy over two unseen datasets, achieving state-of-the-art transcription results on our test split and also on the entire GuitarSet dataset in the zero-shot setting. These strong results add weight to the idea that polyphonic score alignment is an effective way to source training data for AMT systems. In future work we hope to use similar alignment methods to produce a large-scale dataset for guitar music and to further explore applications to other instruments and ensembles.

\vfill
\pagebreak

\bibliographystyle{IEEEbib}
\bibliography{strings,refs}

\end{document}